\pdfoutput=1
\documentclass[a4paper,11pt]{article}
\usepackage{pos}
\usepackage{braket,slashed,mathrsfs,dsfont}
\usepackage{float}
\usepackage{subfiles}

\newcommand{\eg}{\textit{e.g.}~}
\newcommand{\ie}{\textit{i.e.}~}

\renewcommand{\bar}{\overline}

\renewcommand{\epsilon}{\varepsilon}

\newcommand{\QCD}{\mathrm{QCD}}

\renewcommand{\tilde}{\widetilde}

\newcommand{\abs}[1]{\vert{#1} \vert}

\newcommand{\gzero}[1][]{\Gamma_{#1}^\text{tree}} 

\newcommand{\inclrate}[2]{\Gamma( #1^+ \rightarrow #2^+\nu_{#2}[\gamma])}
\newcommand{\zmob}{ZM\"obius\,}

\newcommand{\iso}[1]{\bar{#1}} 
\newcommand{\chipt}{\chi\text{PT}}

\newcommand{\order}{\mathcal{O}}

\newcommand{\pe}[1][]{\partial_{e_{#1}}}
\newcommand{\pmq}{\partial_{m_q}}

\newcommand{\decconst}{f}
\newcommand{\fratio}{\frac{\decconst_K}{\decconst_\pi}} 
\newcommand{\vratio}{\frac{\vert V_{us}\vert}{\vert V_{ud}\vert}} 
\newcommand{\vratiosqrd}{\frac{\vert V_{us}\vert^2}{\vert V_{ud}\vert^2}} 

\newcommand{\dR}{\delta R_{K\pi}}
\newcommand{\dr}[1][]{\delta R_{#1}}

\newcommand{\vus}{V_{us}}
\newcommand{\vud}{V_{ud}}
\newcommand{\Amp}{\mathcal{M}}

\newcommand{\aInv}{a^{-1}}

\title{Near-Physical Point Lattice Calculation of Isospin-Breaking Corrections to $K_{\ell2}/\pi_{\ell2}$}
\ShortTitle{Isospin-Breaking Corrections to $K_{\ell2}/\pi_{\ell2}$}

\author*[a]{Andrew Zhen Ning Yong}
\author[a,b]{Peter Boyle}
\author[a]{Matteo Di Carlo}
\author[a]{Felix Erben}
\author[a]{Vera G\"ulpers}
\author[a]{Maxwell T. Hansen}
\author[a]{Tim Harris}
\author[c]{Nils Hermansson-Truedsson}
\author[a]{Raoul Hodgson}
\author[d,e]{Andreas J\"uttner}
\author[a]{Antonin Portelli}
\author[a]{James Richings}

\affiliation[a]{Department of Physics \& Astronomy, University of Edinburgh,\\
Peter Guthrie Tait Road, King's Buildings, Edinburgh, United Kingdom}
\affiliation[b]{Physics Department, Brookhaven National Laboratory, Upton, NY, USA}
\affiliation[c]{Albert Einstein Center for Fundamental Physics, Institute for Theoretical Physics, Universität Bern, Sidlerstrasse 5, 3012 Bern, Switzerland}
\affiliation[d]{School of Physics and Astronomy, University of Southampton, University Road, Southampton SO17 1BJ, United Kingdom}
\affiliation[e]{CERN, Physics Department, 1211 Geneva 23, Switzerland}
\emailAdd{andrew.yong@ed.ac.uk}

\abstract{In recent years, lattice determinations of non-perturbative quantities such as $f_K$ and $f_\pi$, which are relevant for $V_{us}$ and $V_{ud}$, have reached an impressive precision of $\mathcal{O}(1\%)$ or better. To make further progress, electromagnetic and strong isospin breaking effects must be included in lattice QCD simulations.

We present the status of the RBC/UKQCD lattice calculation of isospin-breaking corrections to light meson leptonic decays. This computation is performed in a (2+1)-flavor QCD simulation using Domain Wall Fermions with near-physical quark masses. The isospin-breaking effects are implemented via a perturbative expansion of the action in $\alpha$ and $(m_u-m_d)$. In this calculation, we work in the electro-quenched approximation and the photons are implemented in the Feynman gauge and $\text{QED}_\text{L}$ formulation.}

\FullConference{%
 The 38th International Symposium on Lattice Field Theory, LATTICE2021
  26th-30th July, 2021
  Zoom/Gather@Massachusetts Institute of Technology
}

\begin{document}
\maketitle

\section{Introduction}

One of the ongoing efforts to search for new physics beyond the Standard Model (SM) of particle physics is to test the unitarity of the Cabibbo-Kobayashi-Maskawa (CKM) matrix. Since all SM and potential beyond-the-SM particles contribute to hadronic decays via virtual corrections, any deviations from the theoretical expectation of unitarity may hint at inconsistencies between Nature and the SM in the flavor sector. In 2020, the PDG \cite{Zyla:2020zbs} reports the following tension with unity in the first row of the CKM matrix:

\begin{equation}
  \vert V_{ud} \vert^2 + \vert V_{us} \vert^2 + \vert V_{ub} \vert^2 = 0.9985 \pm 0.0005.
  \label{eq: CKM first row}
\end{equation}
The determination of CKM matrix elements is an endeavour involving the joint effort of precise experimental measurements and predictions from theory. One of the experimental avenues of interest is the leptonic decay channels of light pseudoscalars, $P^\pm=K^\pm,\pi^\pm$. The tree-level expression is 

\begin{equation}
  \gzero[P]= \frac{G^2_F}{8\pi} M_{P^+}m^2_{\ell^+}\left(1-\frac{m^2_{\ell^+}}{M^2_{P^+}}\right)^2\abs{\decconst_P}^2\abs{V_{qq'}}^2,
  \label{eq: tree-level rate}
\end{equation}
where $G_F$ is the Fermi constant and $V_{qq'}$ is the CKM matrix element between quark flavors $q$ and $q'$. The pseudoscalar decay constant, $\decconst_P$, is a quantity in the isosymmetric theory, where $\alpha=0$ and $\delta m \equiv m_u-m_d = 0$. We note that since the decay constant encapsulates the non-perturbative effects of QCD, it is evaluated numerically using lattice QCD methods. Traditionally, these calculations were performed in the $\alpha=0,\delta m = 0$ regime. However, since recent lattice determinations of $\decconst_K$ and $\decconst_\pi$ have attained percent-level precision \cite{Aoki:2021kgd}, further progress will necessitate the inclusion of isospin-breaking (IB) effects since $\alpha\sim \frac{\delta m}{\Lambda_\QCD}\sim 1\%$.

Due to the experimental challenge in distinguishing between final states with or without a soft photon, only the inclusive rates are measured in the case of pions and kaons. For the extraction of $V_{qq'}$ from experimental data, we combine the tree-level expression with its IB corrections with

\begin{equation}
  \inclrate{P}{\ell} = \gzero[P](1+\delta R_P) + \order(\alpha^2,\delta m^2, \alpha\delta m)
\end{equation}
where $\delta R_P$ contains the leading order IB contributions to the tree-level width.

In this proceeding, we report on our determination of $\vratio$, which can be extracted from a ratio of muonic inclusive rates via 

\begin{equation}
  \vratiosqrd = \frac{\inclrate{K}{\mu}}{\inclrate{\pi}{\mu}} \frac{M_{K^+}}{M_{\pi^+}}\frac{M_{\pi^+}^2-m_{\mu^+}^2}{M_{K^+}^2-m_{\mu^+}^2} \mathcal{F}^{-2},
  \label{eq: vusvud2}
\end{equation}
where
\begin{equation}
  \mathcal{F} = \fratio \sqrt{1+\dR}
  \label{eq: scheme-independent F}
\end{equation}
and $\dR\equiv \dr[K] - \dr[\pi]$. By using experimental measurements of the inclusive rate (\ie branching ratios and mean lifetimes) and masses, the term $\dR$ encapsulates the ratio of IB correction to the hadronic decay amplitudes. The lattice determination of $\dR$ is thus the main focus of this calculation.

\section{Calculation strategy}

The methodology of separating the calculation of real and virtual IB corrections, along with the regulating and removal of finite volume effects (FVE) from the amplitudes computed on the lattice,  was first proposed in \cite{Carrasco:2015xwa}. Following this, a calculation of $\dR$ and $\vratio$ was accomplished in \cite{Giusti:2017dwk, DiCarlo:2019thl}. Here, we adopt a similar strategy and write the inclusive rate as  

\begin{equation}
  \begin{split}
    \inclrate{P}{\mu}&= \lim_{L\rightarrow \infty}(\Gamma_0(L) - \Gamma^{(2)}_0(L)) 
    +  \lim_{\lambda \rightarrow 0} (\Gamma^\text{univ}_0(\lambda) + \Gamma_1(\lambda,\Delta E_\gamma)),
  \end{split}
  \label{eq: Ed-Bern strategy}
\end{equation}
where the subscript denotes the number of real photons in the final state. The virtual corrections in $\Gamma_0$ are evaluated with numerical simulations since all momentum modes of the photon are involved in the interaction with the initial hadron. Thus, the lattice box size, $L$, is a natural choice of IR regulator. To remove the FVE of the lattice calculation, we introduced

\begin{equation}
  \Gamma^{(n)}_0(L) = \Gamma^\text{tree}_P\left(1 + 2 \frac{\alpha}{4\pi}Y^{(n)}_P(L)\right) + \mathcal{O}\left(\frac{1}{L^{n+1}}\right).
\end{equation}
The $\order(L^{-1})$ FVE's were calculated in \cite{PhysRevD.95.034504}, and the $\order(L^{-2})$ structure-dependent FVE's in \cite{DiCarlo:2021apt}. Thus, the residual FVE coming from our lattice calculation now begins at $\Gamma_0(L)-\Gamma^{(2)}_0(L) \sim \order \left(L^{-3}\right)$.

For the real photon contribution, we implement the analytic approach in \cite{Carrasco:2015xwa}. In Equation \eqref{eq: Ed-Bern strategy}, 

\begin{equation}
  \lim_{\lambda \rightarrow 0} (\Gamma^\text{univ}_0(\lambda) + \Gamma_1(\lambda,\Delta E_\gamma)) = \Gamma^\text{tree}_P\left(1 + \frac{\alpha}{4\pi}\delta\Gamma_{1,P}(\Delta E)\right),
\end{equation}
where $\Delta E$ is an energy threshold, below which the photon is sufficiently soft that it treats the initial hadron as a point-like particle. Here, for the IR regulator we use a fictitious photon mass, $\lambda$. Additionally, an intermediate `universal' term, $\Gamma^\text{univ}_0$, is introduced to ensure the IR divergences in Equation \eqref{eq: Ed-Bern strategy} cancel numerically. 
In the following, we will discuss in detail the contributions going into $\Gamma_0(L)$.

\section{Leptonic matrix elements from Euclidean correlation functions}

\begin{figure}[H]
  \centering
  \includegraphics[width=0.65\textwidth]{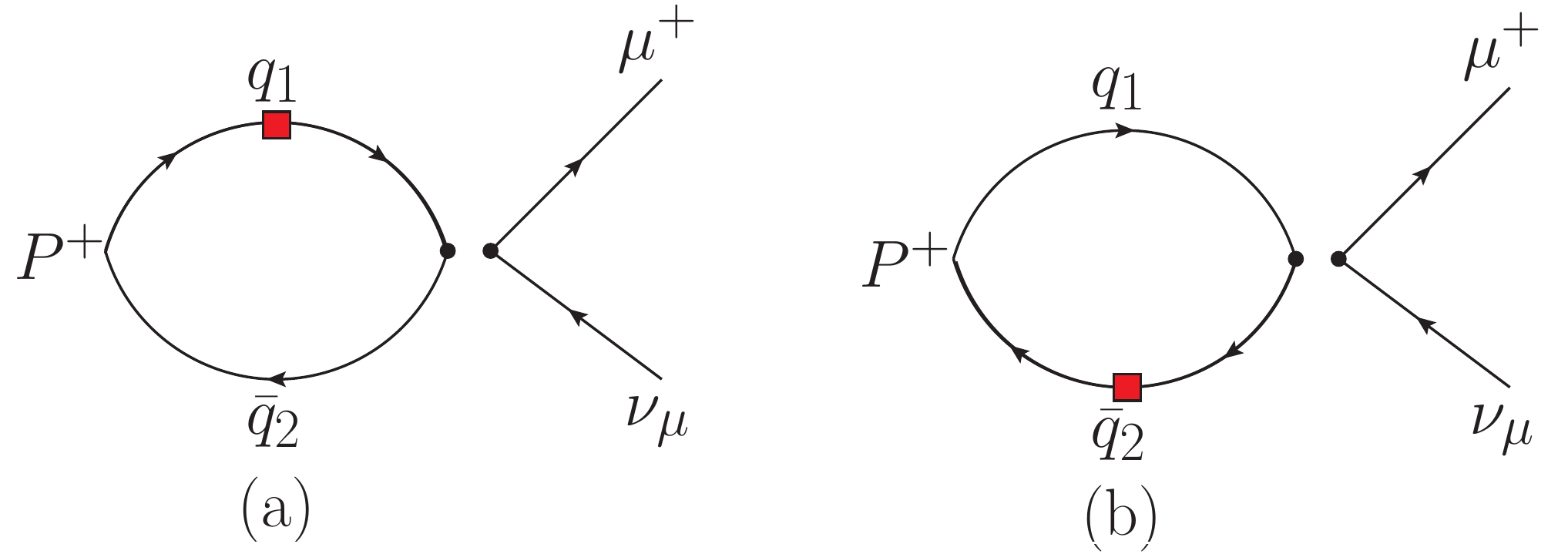}
  \caption{Feynman diagrams of scalar insertions on quark legs (marked with red boxes).}
  \label{fig: SIB diagrams}
\end{figure}

\begin{figure}[H]
  \centering
  \includegraphics[width=0.7\textwidth]{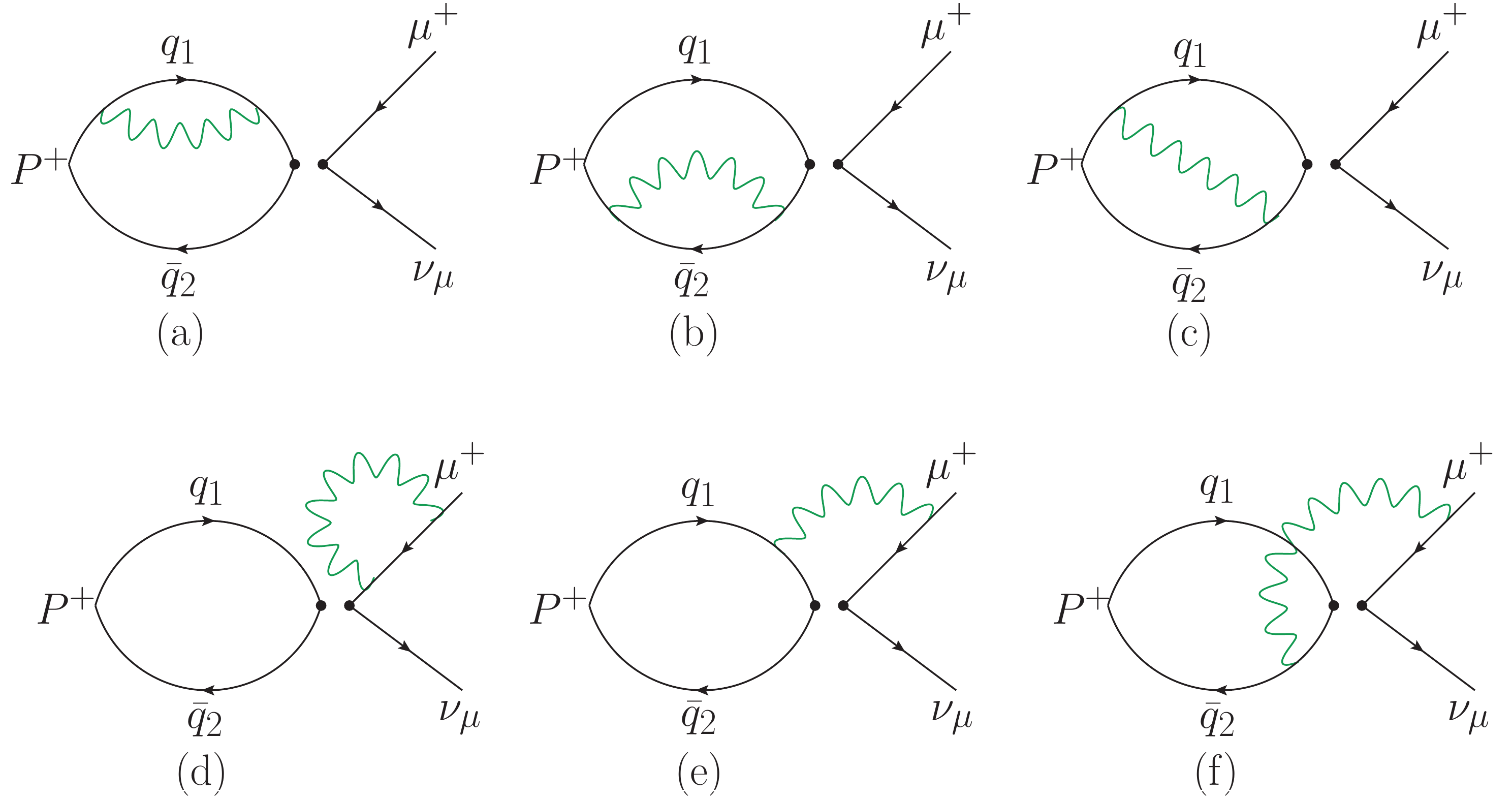}
  \caption{Feynman diagrams of all possible insertions of the electromagnetic current (marked with green squiggle lines) at $\order (\alpha)$. QED interactions with sea quarks are neglected.}
  \label{fig: QED diagrams}
\end{figure}

The 4-fermion operator associated to this decay is
\begin{equation}
   O_W = (\bar{\nu}\gamma^\tau_\text{L}\mu)(\bar{q}_1\gamma^\tau_\text{L}q_2),
\end{equation}
where $\gamma^\tau_\text{L}=\gamma^\tau(\mathds{1}-\gamma^5)$. For a two-body decay, the rate has a simple form:
\begin{equation}
    \Gamma = \mathcal{K}\sum_{r,s}\vert\Amp^{r,s}_P\vert^2
\end{equation}
where $\mathcal{K}$ contains the kinematic factors from the phase space integral and
\begin{equation}
    \Amp^{r,s}_P \equiv \braket{\mu^+,r;\nu_\mu,s|O_W|P^+},
\end{equation}
where $r,s$ are the polarisations of the on-shell final state fermions. In the rest frame of the pseudoscalar, the amplitude in the isosymmetric theory of QCD is 

\begin{equation}
    \iso{\Amp}^{r,s}_P = \left(\bar{u}^r_\nu\Gamma^0_\mathrm{L}v^s_\mu\right)\iso{A}_P,
  \end{equation}
where the axial matrix element is 

\begin{equation}
    \iso{A}_P\equiv \braket{0|\bar{q}_2\Gamma^0_\mathrm{L}q_1|\iso{P}} = \iso{M}_P \decconst_P,
\end{equation}
with $\iso{M}_P$ the pseudoscalar mass in the $\alpha=0$ isosymmetric theory\footnote{Such a theory is scheme-dependent and beyond the scope of the proceedings. Interested readers may refer to \eg \cite{doi:10.1126/science.1257050,Horsley:2015vla,PhysRevD.95.114504,PhysRevD.99.034503}.}. 
At $\order(\alpha,\delta m)$, all the possible virtual IB correction to the isosymmetric amplitude are presented in Figures \ref{fig: SIB diagrams} and \ref{fig: QED diagrams}, which we can write as

\begin{equation}
    \frac{1}{2}\frac{\delta\Gamma_{0,P}}{\iso{\Gamma}_{0,P}} 
    \equiv \frac{\sum_{r,s}\text{Re}\left[\iso{\Amp}^{r,s\,\dag}_P\delta\Amp^{r,s}_P\right]}{\sum_{r,s} \vert\iso{\Amp}^{r,s}_P\vert^2} 
    = \frac{\delta A_P}{\iso{A}_P} - \frac{\delta M_P}{\iso{M}_P} + \frac{\delta\Amp_{P\mu}}{\iso{\Amp}_P},
    \label{eq: virtual IB corrections}
\end{equation}
where $\delta\Amp_{P\mu}$ is the amplitude correction corresponding to diagrams (e,f) in Figure \ref{fig: QED diagrams}. The contributions in Equation \eqref{eq: virtual IB corrections} are extracted from correlation functions generated in lattice simulations. Since $\alpha\sim \frac{\delta m}{\Lambda_\text{QCD}}\sim1\%$ in the low energy regime, the QED and strong IB (SIB) corrections can be treated as a perturbation to our path integral expression \cite{deDivitiis:2013xla}. The full QCD+QED expectation value for some observable $O$ is

\begin{equation}
  \langle O \rangle = \langle O \rangle_0 
  + \sum_q (m_q-\iso{m}_q)\frac{\partial}{\partial m_q} \langle O \rangle\bigg\vert_{m_q=\iso{m}_q}
  + \frac{1}{2!} e^2 \frac{\partial^2}{\partial e^2} \langle O \rangle\bigg\vert_{e=0}
  + \dots ,
\end{equation}
where 
\begin{equation}
  \langle O \rangle = \frac{1}{\mathcal{Z}}\int \mathcal{D}[\psi]\mathcal{D}[\bar{\psi}]\mathcal{D}[U]\mathcal{D}[A]
  \,O[\psi,\bar{\psi},U,A]\, e^{-S_F[\psi,\bar{\psi},U,A]} e^{-S_G[U]} e^{-S_\gamma[A]}
\end{equation}
is the path integral over the usual quark fields, $\psi, \bar{\psi}$; the $SU(3)$ gluonic fields, $U$; and the photon fields, $A$. Here, $\langle O\rangle_0$ is the QCD-only ($m_u=m_d$) expectation value. The SIB and QED corrections are then extracted  from the slopes of the full correlation function, $\langle O \rangle$. Respectively, these derivatives generate correlation functions that contain scalar insertions (Figure \ref{fig: SIB diagrams}) or two insertions of the electromagnetic (EM) current at $\order (\alpha)$ (Figure \ref{fig: QED diagrams}). Together with the subtraction of FVE and the inclusion of the real photon contribution,

\begin{equation}
  \dR 
  = \left(\frac{\delta\Gamma_{0,K}(L)}{\iso{\Gamma}_{0,K}(L)} - \frac{\delta\Gamma_{0,\pi}(L)}{\iso{\Gamma}_{0,\pi}(L)} \right)
  - 2\frac{\alpha}{4\pi}\left( Y^{(2)}_K(L) - Y^{(2)}_\pi(L)\right)
  +\frac{\alpha}{4\pi}\left(\delta\Gamma_{1,K}(\Delta E) - \delta\Gamma_{1,\pi}(\Delta E) \right).
  \label{eq: delta R content}
\end{equation}

In the following, we classify Figure \ref{fig: SIB diagrams}(a,b) and \ref{fig: QED diagrams}(a,b,c) collectively as \textbf{factorisable} diagrams and Figure \ref{fig: QED diagrams}(e,f) as \textbf{non-factorisable} diagrams. We do not consider the lepton self energy term (Figure \ref{fig: QED diagrams}(d)) as this is absorbed in the lepton renormalisation.

\section{Lattice methodology \& implementation}

For this calculation, we generate correlators in a $48^3\times 96$ lattice using near-physical M\"obius Domain Wall Fermions (DWF). The Domain Wall height and the length of the fifth dimension are $M_5=1.8$ and $L_s=24$, respectively \cite{RBC:2014ntl}. To reduce the computational cost of generating near-physical light quark propagators,  we make use of \zmob fermions \cite{Mcglynn:2015uwh} together with the eigenvectors generated by the RBC/UKQCD collaboration for deflation. The 60 QCD gauge configurations used are also generated by the RBC/UKQCD collaboration using the Iwasaki gauge action \cite{IWASAKI1985141}. The sea quark masses are $am^{sea}_l=0.00078, am^{sea}_s=0.0362$. We choose the valence up- and down-quark masses to have the same value as the sea, $am_u=am_d=am^{sea}_l$ and similarly for the valence strange quarks, $am_s=am^{sea}_s$. In this setup, the lattice spacing is $\aInv=1.7295(38)$GeV and the ensemble pion mass is $M_\pi=139.15(36)$ MeV.

The correlators are built from quark propagators with Coulomb gauge-fixed wall sources. As such, we must generate correlators with both wall and point sinks in order to extract the axial matrix element. The implementation of QED on our lattice simulation is as follows: we remove the photon's spatial zero mode with the $\text{QED}_\text{L}$ formalism. As in the case of the scalar current, we sequentially insert a local EM current in Feynman gauge to obtain a sequential propagator with an $\slashed A$ insertion. The correlators built from these propagators are then renormalised by appropriate factors of $Z_V$ \cite{RBCUKQCD:2015joy}. To build the hadron-leptonic correlators corresponding to diagrams (e,f) in Figure \ref{fig: QED diagrams}, we include the charged lepton on the lattice. This is done by generating muon propagators with a free DWF action, using an input mass such that the pole mass of our DWF propagator matches with the experimentally measured value. The propagator is given twisted boundary conditions to conserve 4-momentum of this decay. The neutrino is a spectator fermion in this whole process, so we choose to omit it in the lattice simulation and include it in the analysis stage. We put the pseudoscalar interpolator at the origin and insert the 4-fermion operator on every timeslice, $t_H$, with the muon source-sink separation fixed at, $t_\mu-t_H = 12,16,\dots,36,40$.

In this calculation, we omit SIB contributions coming from sea quarks. We are also working in the electro-quenched approximation of QED - treating the sea quarks as electrically neutral.

\section{Physical predictions from lattice calculations}

Since the two sources of isospin-breaking are $\order(1\%)$ effects, we can perform a linear expansion about the physical point and treat these effects as shifts from the isosymmetric point, where $\alpha=0, \delta m = 0$. Let $X$ be the observable of interest (\eg hadronic mass). Then, at $\order(\alpha,\delta m)$,

\begin{equation}
    X = \iso{X} + \sum_q \Delta m_q \pmq X\vert_{m_f=\iso{m}_f} + \frac{1}{2!}e^2\partial_{e^2}X\vert_{e=0} + \order(\alpha^2,(\Delta m_q)^2, \alpha\Delta m).
    \label{eq: linear expansion}
\end{equation}
where $\partial_{y^n}X = \partial^n X/\partial y^n$. Working to this order, we can set the electric charge in terms of the fine structure constant in the Thomson limit, $e^2=4\pi\alpha_\text{EM}$\footnote{$\alpha_\text{EM} =  7.2973525693(11) \times 10^{-3}$\cite{RevModPhys.93.025010}}. The set of $\Delta m_q$ are the isosymmetric-to-physical point bare quark mass shifts.
For a theory of QCD with 3 flavors, we can solve for the three $\Delta m_q$'s by imposing the following mass ratios:

\begin{equation}
    \frac{\left(aM_P\right)^2}{\left(aM_{\Omega^-}\right)^2} = \frac{\left(M^{exp}_P\right)^2}{\left(M^{exp}_{\Omega^-}\right)^2},
\end{equation}
where we choose $P = \pi^+,K^+,K^0$. Scale setting can also be done by considering the following dimensionful constraint:

\begin{equation}
    a = \frac{aM_{\Omega^-}}{M^{exp}_{\Omega^-}},
\end{equation}
where we chose the omega baryon owing to the clean signal from lattice simulations.

An additional step is required of our setup. Namely, we are not simulating at but close to the desired isosymmetric point and thus we must correct for this mismatch before calculating the IB corrections. Here, we note that since there is no natural phenomena which interacts exclusively with the strong nuclear force, this unphysical definition of the isosymmetric point will depend on the separation scheme we prescribe to our calculation. To that end, consider the following mesonic quantities,
\begin{equation}
    \begin{split}
        M^2_{ud}    &= \frac{1}{2}\left(M^2_{\bar{u}u} + M^2_{\bar{d}d}\right) \approx 2Bm_{ud}+ \dots \\
        \Delta M^2  &= M^2_{\bar{u}u} - M^2_{\bar{d}d} \approx 2B(m_u-m_d)+ \dots \\
        M^2_{K\chi} &= \frac{1}{2}\left( M^2_{K^+} + M^2_{K^0} - M^2_{\pi^+} \right) \approx 2Bm_s + \dots
    \end{split}
\end{equation}
where $\bar{q}q$ are neutral pseudoscalars made from connected-only propagators. Here, we have used the fact that, at leading order partially-quenched $\chipt$ \cite{PhysRevLett.111.252001, Bijnens:2006mk}, these squared masses are proportional to the quark masses we are interested in, with $B$ the chiral condensate. Thus, we can tune our setup to the isosymmetric point by setting $\alpha = 0$ in Equation \eqref{eq: linear expansion} and determine another set of bare quark mass shifts, $\{\Delta m'_q\}$, with the following constraints: 

\begin{equation}
    M^2_{ud} = \left(M^{exp}_{\pi^0}\right)^2, \quad 
    \Delta M^2 = 0, \quad 
    M^2_{K\chi} = \frac{1}{2}\left( \left(M^{exp}_{K^+}\right)^2 + \left(M^{exp}_{K^0}\right)^2 - \left(M^{exp}_{\pi^+}\right)^2 \right).
\end{equation}
Most notably, by fixing $\Delta M^2 = 0$, we emulate the constraint in which $\delta m = 0$.

\section{Constructing the amplitude correction}

\subsection{Factorisable IB correction}
\begin{figure}[h]
    \centering
    \includegraphics[width=0.8\textwidth]{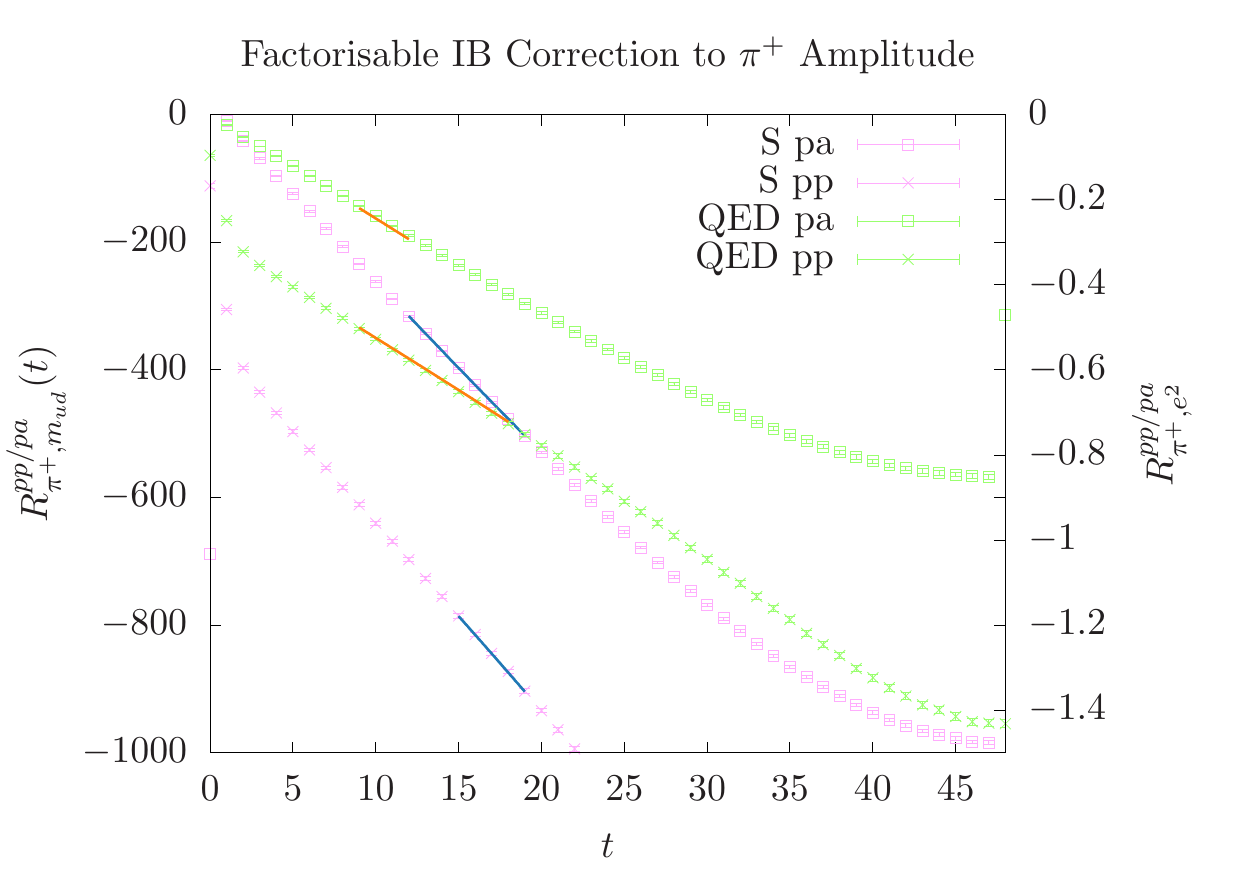}
    \caption{A combined fit of all correlator ratios defined in Equation \eqref{eq: factorisable IB correlator}. The pink points are the $m_{ud}$ scalar insertion data and should be read with the left $y$-axis, while the green points are the QED data and should be read with the right $y$-axis. The blue and orange lines are the fits to the scalar insertion and QED data, respectively. The error bands are not visible. For this combined fit, $\chi^2/dof=70/52=1.35$, $p-$value$=0.099$.}
    \label{fig: factorisable combined fit}
  \end{figure}

The extraction of IB correction to the amplitude from correlators corresponding to Figure \ref{fig: SIB diagrams}(a,b) and Figure \ref{fig: QED diagrams}(a,b,c) proceeds in a manner analogous to the standard two-pt analysis, \ie we need the pseudoscalar ($p$) and axial-pseudoscalar ($a$) interpolators:

\begin{equation}
  \phi_p=\bar{q}_2\gamma^5q_1 \quad\text{and}\quad \phi_{a} = \bar{q}_2\gamma^0\gamma^5q_1.
\end{equation}
Excluding backward propagating effects, the QCD-only $pp/pa-$correlator is 

\begin{equation}
    \iso{C}^{pj}_P(t) \equiv \int d^3x \braket{0|T\left\{\phi_{j}(\vec{x},t)\phi^\dag_p(0)\right\}|0}e^{-i\vec{k}\cdot\vec{x}} = c^{pj}_P e^{-\iso{M}_Pt},
    \label{eq: factorisable tree correlator}
\end{equation}
with $c^{pj}_P =\braket{0|\phi_j|P}\braket{P|\phi^\dag_p|0}/2\iso{M}_P.$ For $pp/pa-$correlators containing either electromagnetic or scalar currents, we have 

\begin{equation}
    \partial_{g_k}C^{pj}_P(t) = \left(\partial_{g_k}c^{pj}_P + c^{pj}_P\partial_{g_k}\iso{M}_Pt\right) 
    e^{-(\iso{M}_P+\Delta g_k\partial_{g_k}\iso{M}_P)t},
    \label{eq: factorisable IB correlator}
\end{equation}
where $g_k=\{ e^2, m_l, m_s\}$ and $\Delta g_k = g_k-\iso{g}_k$. Taking the ratio of Equation \eqref{eq: factorisable IB correlator} and \eqref{eq: factorisable tree correlator}, we have 

\begin{equation}
    R^{pj}_{P,g_k}(t) = \frac{\partial_{g_k}c^{pj}_P}{\iso{c}^{pj}_P} - \partial_{g_k}\iso{M}_Pt.
\end{equation}
Since the QED and SIB correlator ratios share a common mass parameter, $\iso{M}_P$, we include $R^{pj}_{P,g_k}$ for all $j$ and $g_k$ in a combined fit. In Figure \ref{fig: factorisable combined fit}, we present a preliminary result of this combined fit.

\subsection{Non-factorisable IB correction}

\begin{figure}[h]
    \centering
    \includegraphics[width=1\textwidth]{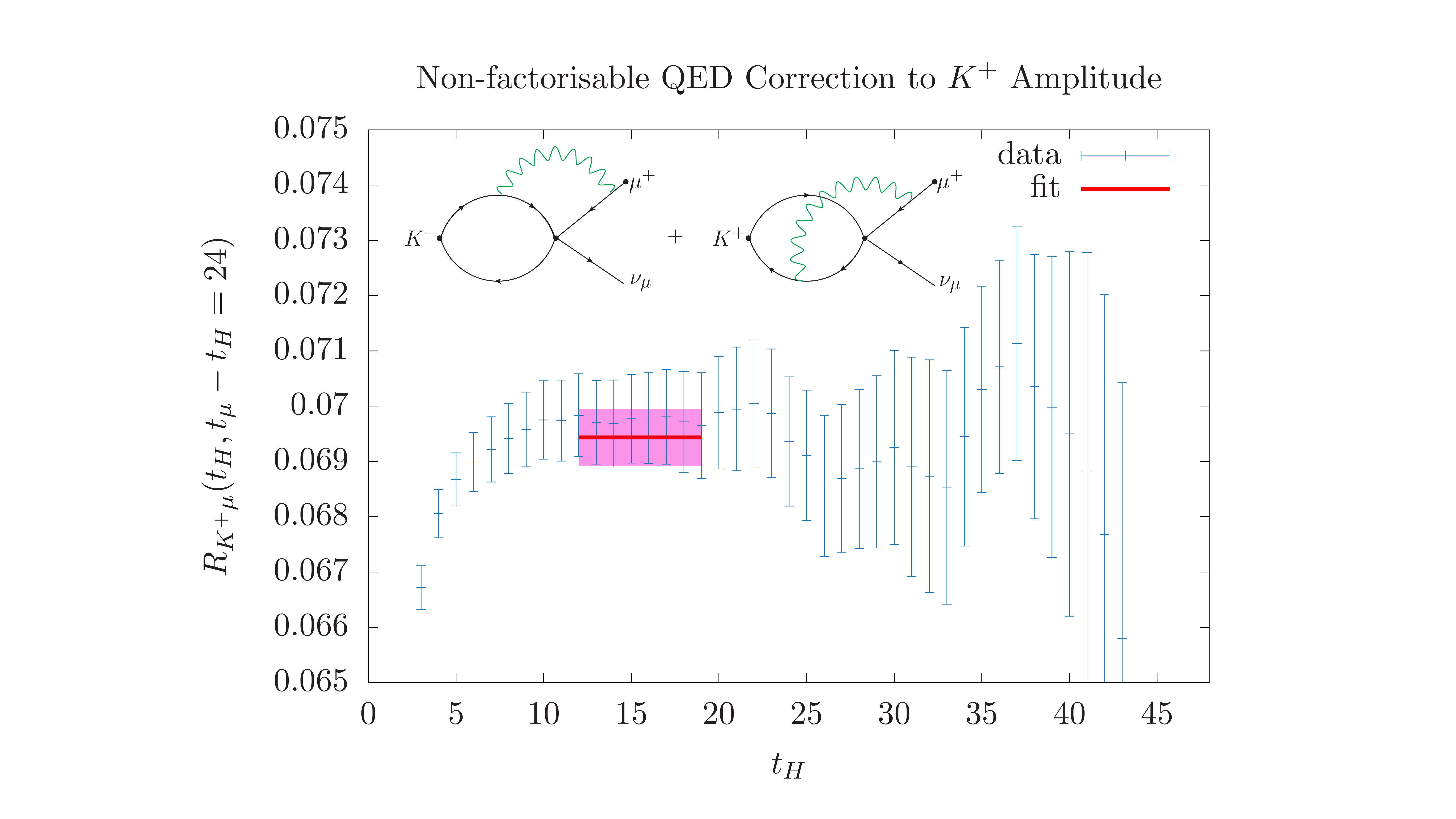}
    \caption{The non-factorisable correlator ratio defined in Equation \eqref{eq: nonfactorisable IB ratio}. The red line and pink error band is the result of a combined fit of the hadron-leptonic correlators, in which only $t_\mu-t_H=24$ is shown here. For this combined fit, $\chi^2/dof=29/31=0.94$, $p-$value$=0.812$. }
    \label{fig: nonfactorisable combined fit}
  \end{figure}

The hadron-leptonic correlators on the lattice are generated with a modified 4-fermion operator:
\begin{equation}
    \tilde{O}_{W,\beta} = (\gamma^\tau_\text{L}\mu)_\beta(\bar{q}_1\gamma^\tau_\text{L}q_2),
\end{equation}
where $\beta$ is an open spinor index. The spectral representation of these novel correlators is
\begin{equation}
    \begin{split}
        \iso{C}_P(t_H,t_\mu-t_H)_{\beta_1\beta_2} &\equiv 
        \int\prod_{j=\mu,H} d^3x_j\,e^{-i\vec{p}_j\cdot\vec{x}_j}\braket{0|\left(\bar{\mu}(\vec{x}_\mu,t_\mu)\Gamma^0_\text{L}\right)\tilde{O}_W(\vec{x}_H,t_H)\phi^\dag(0)|0}_{\beta_1\beta_2} \\
        &= -\frac{\braket{P|\phi^\dag_P|0}}{4E_\mu \iso{M}_P}
        \left[ \iso{\Amp}_P\cdot (\slashed p_\mu - im_\mu) \right]_{\beta_1\beta_2}
        e^{-\iso{M}_Pt_H} e^{-E_\mu(t_\mu-t_H)}
    \end{split}
\end{equation}
where $\beta_{1,2}$ are spinor indices. The correlator containing electromagnetic currents has an analogous functional form. In order to extract the non-factorisable amplitude correction, we saturate the spinor indices by including the missing neutrino leg, giving us a trace over the correlator. Then, taking the ratio of the QED and QCD-only hadron-leptonic correlator, we find that at large time separations, 
\begin{equation}
    \begin{split}
        R_{P\mu}(t_H,t_\mu-t_H) &= 
        \frac{\mathrm{Tr}\, \left[ \slashed p_\nu\pe[q]\pe[\mu]C_P(t_H,t_\mu-t_H)\gamma^0_\text{L} \right]}
        {\mathrm{Tr}\, \left[ \slashed p_\nu\iso{C}_P(t_H,t_\mu-t_H)\gamma^0_\text{L} \right]}\\
        &\xrightarrow{t_\mu\gg t_H \gg 0}\frac{\mathrm{Tr}\, \left[ \slashed p_\nu \pe[q]\pe[\mu]\Amp_P(\slashed p_\mu - im_\mu)\gamma^0_\text{L} \right]}
        {\mathrm{Tr}\, \left[ \slashed p_\nu \iso{\Amp}_P(\slashed p_\mu - im_\mu)\gamma^0_\text{L} \right]}.
    \end{split}
    \label{eq: nonfactorisable IB ratio}
\end{equation}
For each pseudoscalar, we have eight of these correlator ratios, corresponding to the different muon source-sink separations, $t_\mu-t_H$. We perform a 2D combined fit in the $t_H$ and $(t_\mu-t_H)$ direction and Figure \ref{fig: nonfactorisable combined fit} shows the result of this fit analysis.

\subsection{Combining lattice and analytic results}

We are now in the position to construct $\dR$ in Equation \eqref{eq: delta R content}, which we recast here for convenience

\begin{equation}
    \dR 
    = \left(\frac{\delta\Gamma_{0,K}(L)}{\iso{\Gamma}_{0,K}(L)} - \frac{\delta\Gamma_{0,\pi}(L)}{\iso{\Gamma}_{0,\pi}(L)} \right)
    - 2\frac{\alpha}{4\pi}\left( Y^{(2)}_K(L) - Y^{(2)}_\pi(L)\right)
    +\frac{\alpha}{4\pi}\left(\delta\Gamma_{1,K}(\Delta E) - \delta\Gamma_{1,\pi}(\Delta E) \right).
\end{equation}
The output of the combined fits discussed in the previous subsections, properly tuned to the isosymmetric point using Equation \eqref{eq: linear expansion} and the appropriate bare quark mass shifts, $\{\Delta m_q\}$, will give the contribution of the first parentheses on the RHS of the above equation. 
Combining with the analytic results from \cite{DiCarlo:2021apt} and \cite{Carrasco:2015xwa}, corresponding to the second and third parentheses, respectively, we obtain $\dR$.

\section{Conclusion \& Outlook}

A high precision test of the unitarity of the CKM matrix is made possible, in part, with recent improvements in lattice simulations. Through the inclusion of IB effects, we are now in the position to predict light CKM matrix elements at percent-level precision or better. In the RBC\&UKQCD collaboration, we have a lattice setup that allows us to extract the amplitude correction, $\dR$, from a near-physical point simulation. At the time of writing, we are estimating the systematics on the prediction of $\vratio$. Indeed, we expect to provide an update on the ratio of CKM matrix elements shortly after the publication of this proceeding. 

Our immediate plan for the leptonic decay sector is as follows: progress is under way to renormalise the weak operator, $O_W$. This will enable us to obtain $\vert\vud\vert$ and $\vert\vus\vert$ and, in turn, provide an update to the unitarity tension seen in Equation \eqref{eq: CKM first row}. Further improvements are possible with an unquenched calculation, \ie including disconnected contributions and QED interactions with sea quarks. Additionally, we envision in the near future a departure from the point-like approximation to a first principle lattice calculation of the real photon contribution. In the long term, it is our objective to further constrain CKM matrix elements by studying semi-leptonic decays, \eg $K^\pm\rightarrow \pi^0\ell^\pm\nu_\ell$.

\acknowledgements

F.E., V.G., R.H., A.P. and A.Z.N.Y. received funding from the European Research Council (ERC) under the European Union's Horizon 2020 research and innovation programme under grant agreement No 757646.
A.P. is additionally supported by grant agreement No 813942.
P.B. has been supported in part by the U.S. Department of Energy, Office of Science, Office of Nuclear Physics under the Contract No. DE-SC-0012704 (BNL).
M.D.C., V.G., M.T.H., T.H. and A.P. are supported in part by UK STFC grant ST/P000630/1. 
M.T.H. is further supported by UK Research and Innovation Future Leader Fellowship MR/T019956/1.
N.H.-T. is funded by the Albert Einstein Center for Fundamental Physics at the University of Bern. 
A.J. acknowledges funding from STFC consolidated grants ST/P000711/1 and ST/T000775/1.
J.R. is supported by DiRAC grants ST/R001006/1 and ST/S003762/1.

This work used the DiRAC Extreme Scaling service at the University of Edinburgh, operated by the Edinburgh Parallel Computing Centre on behalf of the STFC DiRAC HPC Facility (\href{www.dirac.ac.uk}{www.dirac.ac.uk}). This equipment was funded by BEIS capital funding via STFC capital grant ST/R00238X/1 and STFC DiRAC Operations grant ST/R001006/1. DiRAC is part of the National e-Infrastructure.

\bibliographystyle{JHEP}
\bibliography{proceedings}

\end{document}